\magnification=1200

\font\second=cmr7 scaled\magstep1
\font\secondit=cmmi7 scaled\magstep1
\font\secondb=cmbx7 scaled\magstep1

\hsize 5.5truein
\vsize 8.5truein

\hoffset=0.5truein
\nopagenumbers

\overfullrule=0pt

\def\bra#1{{\langle#1\vert}}
\def\ket#1{{\vert#1\rangle}}

\def\sst#1{{\scriptscriptstyle #1}}

\def\mn{{m_\sst{N}}}
\def\mns{{m^2_\sst{N}}}

\def\sst#1{{\scriptscriptstyle #1}}

\def\mks{{m_\sst{K}^2}}

\def\GES{{G_\sst{E}^{(s)}}}

\def\rsstr{{r^2_s}}

\def\PRC#1{{{\it   Phys. Rev.} {\bf C#1} }}
\def\PRD#1{{{\it   Phys. Rev.} {\bf D#1} }}

\def\NPB#1{{{\it   Nucl. Phys.} {\bf B#1} }}

\hfill DOE/ER/40561-348-INT97-00-187

\centerline{}
\centerline{}

\centerline{\bf Nucleon Strangeness: Theory}

\vskip 2pt

\centerline{\second M.J. Ramsey-Musolf}

\vskip 2pt

{
\centerline{\secondit Department$\>\>$ of$\>\>$Physics$,$
$\quad$University$\>\>$of $\>\>$Connecticut}

\centerline{\secondit Storrs$,\>$ CT$\quad  06269\quad$ USA}
\centerline{\second and}
\centerline{\secondit Institute$\>\>$ for$\>\>$ Nuclear$\>\>$ Theory$,$ 
$\quad$University$\>\>$ of$\>\>$ Washington}
\centerline{\secondit Seattle$,\>$ WA$\quad  98195\quad$ USA}

\medskip

{\noindent \second{\secondb Abstract.} The status of theoretical calculations 
of the strange quark vector current form factors of the nucleon is reviewed.}}

\bigskip

The role played by the $q\bar{q}$ sea in the low-energy structure of the
nucleon remains a topic of on-going interest in hadron structure theory.
Of particular interest is the strange quark component of this sea, since
$s\bar{s}$ pairs constitute the lightest pure sea quark degree of freedom
in the nucleon. A study of strange quarks therefore entails no ambiguity
in separating sea from valence quark effects. Moreover, the mass scale
associated with virtual $s\bar{s}$ pair -- $m_s\sim \Lambda_{QCD}$ --
implies that such pairs live for a sufficiently long time and propagate
over sufficiently large distances to produce observable effects when
probed explicity. Such explicit probes of the structure of the 
$s\bar{s}$ sea are currently underway using lepton-nucleon and lepton-nucleus
scattering [1]. These experiments will measure matrix elements of the strange
quark currents: $\bra{N}\bar{s}\gamma_\mu s\ket{N}$ and $\bra{N}\bar{s}
\gamma_\mu\gamma_5 s\ket{N}$. A survey of these experiments will be given
in the following talk by E. Beise, so I will not provide any of the details
here.

Theoretically, the strange quark sea -- and that of virtual $q\bar{q}$
pairs generally -- is interesting because it bears on the question: Why
does the quark model work so well? Indeed, there exists little convincing
evidence for the need to include the sea quarks as an explicit degree
of freedom in accounting for the low-energy properties of hadrons. A 
partial explanation for this situation has been given by Isgur [2], who noted
that in the adiabatic approximation, the effect of virtual $q\bar{q}$
pairs is effectively to renormalize the string tension. In the quark
model, the latter is determined phenomenologically, rather than via
an {\it ab initio} calculation. Consequently, the effect of $q\bar{q}$ pairs
is already included, and no explicit signature of sea quarks appears in
the hadronic spectrum. An alternative explanation has been suggested by
Kaplan and Manohar [3], who propose that the effect of the quark and gluon
sea is to renormalize the current quarks of QCD into the constituent quarks
of the quark model. In either case, the presence of the $s\bar{s}$ sea
is \lq\lq hidden" in observables that do not carry the quantum numbers
of strangeness. What makes the lepton probes mentioned above so interesting
is that they hope to uncover this \lq\lq hidden" degree of freedom and
provide new insight into the low-energy structure of the $q\bar{q}$ sea.

Ideally, one would like to have some reliable theoretical predictions for
the scale and sign of strange quark matrix elements in order to compare
with measurements -- or, at least, have in hand a useful theoretical
framework for interpreting them. In fact, the last several years has 
witnessed a plethora of theoretical predictions for the form factors
which parameterize the strangeness matrix elements. These predictions
fall under three general headings: (a) first principles calculations
using lattice QCD [4]; (b) hadronic models [5]; (c) effective hadronic 
[6-8] theory. At present a few results have been obtained from the lattice 
in the quenched approximation for the strangeness \lq\lq magnetic moment"
and axial charge [4]. However, a refined lattice calculation appears to
be a long way off, and even when it is achieved, one may still require
other theoretical tools in order to understand the mechanisms responsible
for the scale and sign of the lattice results. QCD-inspired hadronic models 
are an appealing approach, as they generally emphasize a particular
hadronic mechanism and provide a mental picture of strange quark dynamics.
Nevertheless, the connection between a given model and QCD is not
rigorous, and one must ask whether a given model result is adequately
reflective of the full range of strong interaction dynamics. The ambiguities
associated with hadronic models are reflected in the rather broad range
of theoretical predictions for the strange quark form factors.

An alternative approach, which I wish to emphasize in the remainder of
this talk, is the use of effective hadronic theory. The most popular
recent version of hadronic effective theory is chiral perturbation
theory (CHPT). While CHPT has enjoyed considerable success in accounting
for a variety of low-energy hadronic properties, it is rather limited
in the case of nucleon strangeness [6]. The reason has to do with the
symmetry structure of the strangeness currents. Focusing on the vector
current, one has

$$
\bar{s}\gamma_\mu s = J_\mu^{\rm baryon}-2J_\mu^{\rm em}(I=0)\ \ \ ,
\eqno(1)
$$
where $J_\mu^{\rm baryon}$ is the baryon number current (an SU(3)-singlet)
and $J_\mu^{\rm em}(I=0)$ is the isoscalar electromagnetic current. 
The structure in Eq. (1) implies that the \lq\lq low-energy constants"
or chiral counterterms relevant to the strangeness vector current matrix
elements cannot be determined from existing data using symmetry. While
there exists sufficent data in the electromagnetic sector, there does
not exist the corresponding data in the SU(3)-singlet, or baryon number,
sector. Hence, the pieces of the chiral counterterms that depend on
baryon number are unknown at present. In fact, measurements of strangeness
matrix elements effectively determines these SU(3)-singlet counterterms.

An effective hadronic approach which is similar in spirit to CHPT but
more fruitful in the case of strangeness is that of dispersion relations
(DR's). Like CHPT, which relies on chiral symmetry to relate existing
data to the quantities one would like to predict, DR's rely on analaticity
and causality to relate existing scattering data to the strangeness form
factors. To illustrate, consider the mean square \lq\lq strangeness radius"
of the nucleon, $\rsstr$, defined as the slope of the strangeness electric
form factor at the photon point. The DR for this quantity is
$$
\rsstr={6\over\pi}\int_{t_0}^\infty\ dt {\hbox{Im} \GES(t)\over t^2}
\ \ \ , \eqno(2)
$$
where $t=q^2$ and $t_0$ is a threshold to be discussed shortly. Various
techniques from field theory allow one to relate $\hbox{Im} \GES(t)$ --
called the spectral function -- to amplitudes for a nucleon and anti-nucleon
to annihilate into appropriate physical states $\ket{n}$ having mass
$\sqrt{t_0}$. The lightest such state which can contribute is $\ket{3\pi}$,
which model-builders usually neglect because it contains no valence
strange quarks. 

The lightest state containing valence strange quarks is
$\ket{K\bar{K}}$. Its contribution to the spectral function is [7,8]
$$
\hbox{Im} \GES(t)^{K\bar{K}} =  \hbox{Re} 
\left\{
\left( {\sqrt{t-4\mks}\over 8\mn} \right) 
b_1^{1/2,\ 1/2}(t) F_K^{(s)}(t)^{*}\right\}\ \ \ ,\eqno(3)
$$
where $b_1^{1/2,\ 1/2}$ is the one of the two $N\bar{N}\to K\bar{K}$ $J=1$
partial waves and $F_K^{(s)}$ is the kaon strangeness form factor.
The former can be determined by performing fits to $K^{+} N$ scattering
and analytically continuing to the kinematic region needed for the
integral in Eq. (2). The kaon strangeness form factor can be obtained
by applying straightforward flavor rotation arguments to the kaon
electromagnetic form factor, which is known from $e^+e^-$ annihilation
data. 

H.-W. Hammer and I recently completed a calculation of the $K\bar{K}$
contribution to $\rsstr$ using this technique [8]. The corresponding
contribution to the integrand of Eq. (2) is plotted below. Two curves
are shown in the un-physical regime ($4 m_K^s\leq t\leq 4\mns$).
The dot-dashed curve shows the result obtained using the
analytically continued $K^+ N$ amplitudes and realistic kaon strangeness
form factor. The solid curve shows the same quantity computed at second
order in the strong coupling, $g$ ({\it i.e.}, one-loop). The latter is
representative of what enters a variety of model calculations.  The dashed
curve in the physical region ($t\geq 4\mns$) results from the imposition
of the unitarity bound on $b_1^{1/2,\ 1/2}$. 

Several features in this plot are worth noting. First, the existing data
for $K^+N$ partial waves only permit an analytic continuation of the
amplitudes to about $t\approx 8 m_K^2$ with any reliability. However,
the kaon strangeness form factor falls rapidly below unity for 
$t> 1.4\ (\hbox{GeV}/c)^2$. Consequently, any contribution to the dispersion
integral for this region is negligible. Moreover, both the strangeness
form factor and $b_1^{1/2,\ 1/2}$ display a pronounced peak near $t\approx
1\ (\hbox{GeV}/c)^2$, presumably arising from the $\phi(1020)$ resonance.
This structure differs markedly from the content of a typical one-loop
model calculation, which contains none of the resonance structure near
the threshold, $t_0$ and violates the unitarity bound rather drastically
in the physical region [7]. The latter feature reflects the absence of higher
order rescattering corrections responsible for bringing the partial wave
in line with the unitarity bound. 

\vskip 3.0 true in

The message from the figure is that simple model calculations which 
involve truncations in the coupling constant (usually at second order)
omit the physics which governs the various hadronic contributions to the
strangeness form factors. Numerically, inclusion of these higher-order
rescattering and resonance effects enhances this particular contribution
($K\bar{K}$) to $\rsstr$ by about a factor of three over the order
$g^2$ prediction. Recently, Isgur and Geiger have shown that models which
make another truncation -- one involving the intermediate state mass --
can be similarly misleading [9]. Including the full spectrum of hadronic
intermediate states, at least those involving valence strange quarks, can
dramatically alter the prediction obtained from just the lightest state. 
Although the Isgur and Geiger results were obtained from an ${\cal O}
(g^2)$ quark model calculation, their results strongly imply that realistic
treatments of higher-mass intermediate states is necessary in order to
obtain a physically realistic picture of the $s\bar{s}$ sea. Just what that
picture looks like, when higher order rescattering corrections, resonance
effects, and higher-mass hadronic states are included, remains to
be seen.

\bigskip
\centerline{\bf References}

\bigskip
\noindent 1. MIT-Bates proposal No. 89-06, R.D. McKeown and D.H.
Beck, spokespersons (1989); MIT-Bates proposal No. 94-11, M. Pitt
and E.J. Beise, spokespersons (1994); TJNAF proposal No. PR-91-017,
D.H. Beck, spokesperson (1991); TJNAF proposal No. PR-91-004, E.J.
Beise, spokesperson (1991); TJNAF proposal No. PR-91-010, M. Finn and
P. Souder, spokespersons (1991); Mainz proposal No. A4/1-93, D.
von Harrach, spokesperson (1993); Los Alamos proposal No. 1173,
W.C. Louis, spokesperson.

\medskip
\noindent 2. P. Geiger and N. Isgur, \PRD{41} (1990) 1595.

\medskip
\noindent 3. D. B. Kaplan and A. Manohar, \NPB{310} (1988) 527.

\medskip
\noindent 4. K.-F. Liu, University of Kentucky Report No. UK/95-11
(1995) and references therein; D. B. Leinweber, \PRD{53} (1996) 5115.

\medskip
\noindent 5. See Refs [30-42] of Ref. [6] below.

\medskip 
\noindent 6. M. J. Ramsey-Musolf and H. Ito, \PRC{55} (1997) 3066.

\medskip
\noindent 7. M. J. Musolf, H.-W. Hammer, and D. Drechsel, \PRD{55}
(1997) 2741.

\medskip 
\noindent 8. M. J. Ramsey-Musolf and H.-W. Hammer, INT preprint
No. DOE/ER/40561-323-INT97-00-170 [hep-ph/9705409].

\medskip 
\noindent 9. P. Geiger and N. Isgur, \PRD{55} (1997) 299.

\vfill
\eject
\end